\documentclass[runningheads]{llncs}
\usepackage{makeidx}
\usepackage[T2A]{fontenc}
\usepackage[utf8]{inputenc}
\usepackage[english]{babel}
\usepackage{indentfirst}
\usepackage{amsfonts,amsmath}
\usepackage{color}
\usepackage[final]{graphicx}
\usepackage{epstopdf}
\usepackage[labelsep=period]{caption}
\usepackage[hyphens]{url}

\usepackage{siunitx}

\begin{document}

    \mainmatter

    \title{Multi-Core Processor Scheduling with Respect to Data Bus Bandwidth}

    \author{
        Anton V. Eremeev\inst{1,3}
        %\orcidID{0000-0001-5289-7874}
    \and\\
        Anton A. Malakhov\inst{2}
    \and\\
        Maxim A. Sakhno\inst{1,3}
    \and\\
        Maria Y. Sosnovskaya\inst{1,3}
    }

    \institute{
        Sobolev Institute of Mathematics, Novosibirsk, Russia, \\
        \and
        Intel Corporation, Nizhny Novgorod, Russia, \\
        \and
        Dostoevsky Omsk State University, Omsk, Russia \\
    }

    \authorrunning{A. Eremeev, A. Malakhov, M. Sakhno, and M. Sosnovskaya}

    \maketitle

    \begin{abstract}
        The paper considers the problem of scheduling software modules on a multi-core processor, taking into account the limited bandwidth of the data bus and the precedence constraints. Two problem formulations with different levels of problem-specific detail are suggested and both shown to be NP-hard. A mixed integer linear programming (MILP) model is proposed for the first problem formulation, and a greedy algorithm is developed for the second one. An experimental comparison of the results of the greedy algorithm and the MILP solutions found by CPLEX solver is carried out.

    \end{abstract}

    \section{Introduction} \label{section:introduction}

    The goal of the paper is to investigate resource constraint scheduling problems that arise when developing a program for a multi-core processor. In this case, it is necessary to schedule the execution of software modules on the processor cores, taking into account the restrictions on the data bus bandwidth. The data bus is a part of the system bus that is used to transfer data between computer components, in this particular case, between the CPU and the random access memory~(RAM). Different software modules need different amount of data flow via data bus, therefore in the case of simultaneous execution of several modules, each one of them can take longer time than in the case of single-thread execution. The problem of scheduling software modules on a multi-core processor w.r.t the limited bandwidth of the data bus is important for processor manufacturers and parallel software developing companies, since the more efficiently the data bus is used, the higher the software performance.

From the point of view of scheduling theory, the problem of allocating the software modules to processor cores with respect to the limited data bus bandwidth is similar to the scheduling problems with renewable resources (see e.g.~\cite{Servakh}), but unlike those problems, in our case the resource constraint (data bus bandwidth) does not exclude some infeasible combinations of jobs (software modules) but rather increases their execution times.
%
%     In~\cite{Servakh}, the polynomial solvability of a scheduling problem with limited renewable resources and unit job durations was proved, if the width of the partial order specified on the set of jobs is bounded by a constant. In~\cite{Kovalenko_article}, a dynamic programming algorithm was developed for a scheduling problem with a renewable resource and a variable intensity of its consumption, also it was proved that if the number of machines is bounded by a constant, then the problem is pseudo-polynomially solvable and a polynomial solvable special case is identified.
%
A distinctive feature of our problem is that each job would be processed at different speeds depending on its requirement of the data bus bandwidth and the loading of the data bus by the simultaneous jobs on other cores.

There are a number of approaches to task scheduling with variable processing times in the literature. First of all, in the area of parallel software development for multi-core processors, such problems are usually solved using fast heuristics, which work in the online mode, i.e. the jobs arrive sequentially and only a limited number of jobs is considered in each moment. The task scheduling heuristics proposed in~\cite{Merkel},~\cite{Zhuravlev} and some other works are based on the principle that tasks should be allocated on the CPU cores in a complementary fashion, so that the tasks with most different resource consumption requirements are co-scheduled for simultaneous execution (in~\cite{Merkel},~\cite{Zhuravlev} such resources imply the usage of data bus bandwidth and the cache utilization at different levels).

The tasks scheduling method proposed in~\cite{Jiang} is based on the co-run {\em degradation coefficients}, equal to an increase in the execution time of an application when it shares a cache with a co-runner, relative to running solo. In the case of dual-core CPUs, the threads may be represented as nodes connected by edges, and the weights of the edges are given by the sum of the mutual co-run degradations between
the two threads. Then, under some simplifying assumptions, an optimal schedule may be found by solving a min-weight perfect matching problem. In the case of greater number of cores per CPU, the problem is shown to be NP-hard and several heuristic approximation algorithms are suggested. Although the methodology from~\cite{Jiang} and the corresponding algorithms
would be too expensive to use online, they are acceptable for offline evaluation of the quality of other approaches.

Authors of ~\cite{Xiao} propose a novel fairness-aware thread co-scheduling algorithm based on non-cooperative game to reduce L2 cache misses. The execution time of a thread varies depending on which threads are running on other cores of the same chip, because different thread combinations result in different levels of cache contention. In~\cite{Tian}, the cache on each of ${m}$ chips is shared by ${u}$ cores on the each chip. The execution speed of a job running on a chip depends on what jobs are placed on the same chip. The number of jobs is equal to the total number of cores, all the jobs start at the same time. It is proved that the problem is NP-hard and a series of algorithms is presented to compute or approximate the optimal schedules.

In the production scheduling applications, the problem formulations with variable processing times are also important, e.g.
in \cite{Liu}, a coke production scheduling problem is considered, where jobs influence on the processing time of other jobs due to increased production unit temperature. The authors of \cite{Liu} construct an integer programming model to minimize
the makespan, propose several heuristics, including a genetic algorithm, and compare their performance.

In the scheduling theory, similar problem formulations may be found in the area of scheduling with controllable processing times. The models and methods for the case of preemptive scheduling are surveyed in~\cite{Strusevich}.
In \cite{Shabtay}, the problem of job scheduling on identical parallel machines is considered, where the processing time of jobs is controlled by allocating a non-renewable shared limited resource. It is proved that if job preemptions are allowed, then the problem of minimizing the makespan time is solvable in ${O(n^2)}$ operations.
In the present paper, however, we consider a non-preemptive problem formulation. Since the data bus bandwidth is a renewable resource, we refer to~\cite{Jozefowska} and \cite{Brauner} for the surveys on problems with renewable resources, where the resource allocations may vary over time.
The case of discrete resources is considered in~\cite{Brauner}, and the continuous resources are considered in~\cite{Jozefowska}. The latter paper contains a problem formulation similar to our formulation~F1 considered below, however in \cite{Jozefowska} it is supposed that the amount of resources allocated to each job is limited, but continuous and decided by the scheduler at each moment of time. In our case, the jobs execution speeds are completely defined by the set of co-scheduled jobs on the other cores (or machines in the traditional scheduling terminology).
%
%The authors of~\cite{Gawiejnowicz} consider pairs of so-called {\em isomorphic} scheduling problems, where a `'classical'' scheduling problem with fixed job processing times is considered together with its time-dependent counterpart where the processing times are proportional to the jobs starting times. It is shown that some polynomial algorithms and approximation algorithms for scheduling problems with fixed job processing times may be converted into analogous algorithms for proportional-linear counterparts of the problems.
%
In a recent work~\cite{Althaus18}, the authors focus on assignment of shared continuous resources {\em to the processors}, while the job assignment to processors and the ordering of the jobs is fixed. These are the main differences to the problem considered in our paper. One more difference is that unlike~\cite{Althaus18}, we make a continuous time assumption. The authors of~\cite{Althaus18} show that, even for unit size jobs, finding an optimal solution is NP-hard if the number of processors is a part of the input, however a polynomial-time algorithm for any constant number of processors and unit size jobs exists.

In the present paper, two mathematical problem formulations for
the problem of allocating the software modules to processor cores are proposed with different levels of problem-specific detail and both shown to be NP-hard. A mixed integer linear programming (MILP) model using the concept of event points (see e.g.~\cite{Ierapetritou})
is proposed for the more detailed problem formulation, and a greedy algorithm is developed for the other one. A comparison of the greedy algorithm results and the MILP solutions found by CPLEX solver is carried out.

{
The paper has the following structure. Two problem formulations are proposed in Section~\ref{section:problem_formulation}. NP-hardness of both problem formulations is shown in Section~\ref{section:np_hardness}. A mixed integer linear programming model for the first problem formulation is suggested in Section~\ref{section:milp_model}. The greedy heuristic for the second problem formulation is described in Section~\ref{sec:greedy}.
Methods of real-life input data generation and testing are explained in Section~\ref{section:data_generation}.
The results of computational experiments are presented in Section~\ref{section:computational_experiment}.
Concluding remarks are given in Section~\ref{section:conclusions}.
}

    \section{Problem Formulations} \label{section:problem_formulation}

{
    Informally, our problem is to schedule execution of software modules (jobs) on a number of processor cores, while there is one resource of a renewable type, the bandwidth of the data bus, and the precedence constraints for execution of these modules are given as a partial order on the set of jobs, and the objective is to minimize the makespan. Here we assume that each module creates a uniform data flow through the data bus, so that the amount of information sent by a module through the data bus in both directions (from CPU to RAM and back) is proportional to the fraction of the completed job (i.e. the ratio of the executed elementary operations of a job to the total number of elementary operations in this job). Examples of software modules with (almost) uniform data flow may be the computational routines with multiple repetitions of the same loop or copying large data arrays.
    }

    {\bf Formulation F1.} There are ${m}$ jobs, ${c}$ processor cores. The jobs are performed with no preemption and do not migrate from one core to another during the execution. No more than one job can be performed on a single core.

    For each job ${p}$, ${p = 1, ..., m}$, let ${s_{p}}$ denote its processing time under ideal conditions. Here and below, by {\em ideal conditions} we mean job execution when no other job is performed simultaneously.

We will call {\em a configuration} any set of jobs which may be performed simultaneously on different cores, taking into account the partial order on the set of jobs (a configuration can not contain a pair of jobs where one job precedes another according to the partial order) and the restrictions on the number of cores. Let ${K}$ denote the set of all configurations. Suppose that in zero configuration no job is performed. Clearly, the partial order on the set of jobs also induces a partial order on the set of configurations~${K}$.

Let us call {\em a processing speed} of job ${p}$ in configuration ${k \in K}$ the ratio of the time of full execution of job ${p}$ under ideal conditions to the time of full execution of job ${p}$, if ${p}$ was executed all this time in configuration ${k}$. Throughout each configuration, the speed of all jobs is supposed to be constant, but the processing speed of a job may vary during its execution, depending on the configuration in which it is performed. The configuration can be changed in two cases: the first case is when one of the jobs in the current configuration has completely completed and the second case is when some job(s) is added to the current configuration. If the configuration is changed, the speed of those jobs that are still in progress may change.

    So, for each configuration of ${k \in K}$ we know which jobs it consists of. For each job ${p}$ in configuration ${k}$, the speed of its execution ${v_{pk}}$ is known.

    The problem consists in scheduling the jobs on the processor cores with the minimum makespan (i.e. the time of completion of all jobs).

    Since the number of configurations can be very large (up to ${\sum_{i=0}^{c} {m \choose i} %C_{m}^{i}
    }$, depending on the partial order), the problem formulation~F1 can be simplified by introducing the assumption that the job execution speeds are calculated based on their actual consumption of the data bus. In practice, the job speed depends on a large number of factors such as the number of memory access channels for the processor, the number of processor cache levels and their free volume, the processor frequency and its temperature (depending on the specific processor and related components). Explicit consideration of all these factors is beyond the scope of this paper. Based on practice, we suggest another problem formulation which is based on the jobs usage of data bus bandwidth. This problem formulation can be written as follows.

   {\bf Formulation F2.} There are ${m}$ jobs, ${c}$ processor cores and one renewable resource, the data bus.
    Just like in Formulation~F1, the jobs are performed without preemption or migration from one core to another, and a partial order on jobs is given. It is required to schedule the jobs on the processor cores with the minimum makespan.

    Now we suppose that for each job ${p}$, ${p = 1, ..., m}$, the percentage of data bus consumption ${b_{p}}$ under ideal conditions is known. During the execution of job ${p}$, a smaller percentage of the data bus can be allocated than ${b_{p}}$ if other jobs are simultaneously performed on other cores. Denote by ${z_{pk}}$ the actually allocated percentage of the data bus to job ${p}$ in configuration ${k}$. In practice, the distribution of values $z_{pk}$ among the threads is very hardware-specific and depends on many factors, which we can not afford to take into account (see e.g.~\cite{Nesbit06}). As a simple approximation, we assume that the data bus bandwidth allocation to jobs in any configuration~$k$ may be found by Algorithm~1, described below. The speed ${v_{pk}}$ of execution of job ${p}$ in the configuration ${k}$ is then proportional to the ratio ${z_{pk}}/{b_{p}}.$

    \vspace{5mm}
    \textbf{Algorithm 1.} Calculation of data bus consumption for a given configuration

    \textbf{Step 0.} Put the percentage of the free data bus ${freePercent := 100\%}$ (the entire data bus is free) and set the number of jobs for which the data bus is not allocated, ${jobsCount}$ to be the number of jobs in configuration~${k}$.

    \textbf{Step 1.} While ${jobsCount}$ is not ${0}$, do:

    \indent \indent \textbf{1.1} Calculate a percentage of data bus that can be allocated to each job:

    \indent \indent \indent \indent ${percent := freePercent / jobsCount}.$

    \indent \indent \textbf{1.2} If the configuration has such a job ${p}$ that ${b_{p} < percent}$, then put:

    \indent \indent \indent \indent ${z_{pk} := b_{p}},$

    \indent \indent \indent \indent ${freePercent := freePercent - b_{p}},$

    \indent \indent \indent \indent ${jobsCount := jobsCount - 1}.$

    \indent \indent \indent If no such job is found, then put ${z_{pk} := percent}$ for each remaining job and ${jobsCount := 0}.$

    \textbf{Step 3.} Output computed values $z_{pk}$.\\

This method of capacity allocation is different from the concurrent network flow allocation, well-known in multicommodity flow problems (see e.g.~\cite{SM90}), where the ratio of the flow of each commodity to the predefined flow demand for that commodity must be the same for all commodities. We expect that the capacity allocation represented by Algorithm~1 is more adequate to the case of data bus information flows because in this case a software module has no explicit way to communicate its flow demand to the system.

    \section{Problem Complexity} \label{section:np_hardness}

We will show that the decision versions in both formulations contain an NP-complete special case of {\sc Multiprocessor Scheduling} problem~\cite{GJ} as a special case. Here is a formulation of this problem:

{\em    Given a set ${T}$ of tasks, number ${w \in \mathbb {Z}^{+}}$ of processors, length ${l(t) \in \mathbb{Z}^{+}}$ for each ${t \in T}$, and a deadline ${D \in \mathbb{Z}^{+}}$, is there a ${w}$-processor schedule for ${T}$ that meets the overall deadline ${D}$?}

    In \cite{GJ} it is also proved that the {\sc Multiprocessor Scheduling} problem remains NP-complete in the special case of ${w = 2}$.

    \begin{proposition}
The problem of multi-core processor scheduling with respect to data bus bandwidth is NP-hard for both formulations F1 and F2.
    \end{proposition}

    \textbf{Proof.} The {\sc Multiprocessor Scheduling} problem is a special case of the decision version of the problem of multi-core processor scheduling with respect to data bus bandwidth in Formulation F2 in the special case where: (i)~jobs do not slow each other, (ii)~there is no partial order constraint, and (iii)~the set of tasks ${T}$ is equal to the set of jobs, assuming that the number of processors is the number of cores.

    To prove the NP-hardness of the problem in Formulation F1, put the number of cores equal to ${2}$. In this case, the number of configurations is ${1 + m + \frac{m(m-1)}{2}}$ and, therefore, the input size of the problem in question is limited by a polynomial of the input size of the {\sc Multiprocessor Scheduling} problem. $\Box$
% Thus, the problem of multi-core processor scheduling with respect to data bus bandwidth is NP-hard for the formulations F1 and F2.

    \section{Mixed Integer Linear Programming Model} \label{section:milp_model}

    Consider a mixed integer linear programming (MILP) model for the first problem formulation. We define the concept of an event point similar to that introduced in \cite{Ierapetritou}. In this paper, an event point characterises a time interval in which a single configuration is performed. It is defined by the number of the interval, its duration and the configuration used in it.

   Let ${P = \left\{1,..., m\right\}}$ denote the set of all jobs. The following set of parameters may be computed on the basis of an instance given in Formulation~F1:
\begin{itemize}
\item    ${q_{pk} = 1}$ if and only if job ${p}$ is performed in configuration ${k}$, ${0}$ otherwise.

\item    ${a_{ij} = 1}$ if and only if configuration ${i}$ should run after configuration ${j}$, ${0}$ otherwise.

\item    ${T_{max}}$ is an upper bound on the duration of any configuration at any event point.
\end{itemize}

  Let us denote by ${N = \left\{ 0, 1, 2, ..., e \right\}}$ the set of all event points, where $e$ is the maximal index of event points, and introduce the problem variables:
\begin{itemize}
\item      ${t_{nk}}$ is duration of execution of configuration ${k}$ at the event point ${n}$.

\item      ${d_{nk} = 1}$ if and only if configuration ${k}$ is performed at the event point ${n}$, ${0}$ otherwise. For consistency of the MILP model, we assume that the zero configuration is performed at the zero point of events. Then ${d_{00}=1}$ and ${d_{0k} = 0}$, ${k \in K}$.

\item      ${y_{pn} = 1}$ if and only if job ${p}$ started at the event point ${n}$, ${0}$ otherwise.
\end{itemize}

    Then the MILP model can be written as follows:

    \begin{equation}\label{eqn:milp1_2}
        \min \sum_{n \in N} \sum_{k \in K} t_{nk},
    \end{equation}

    \begin{equation}\label{eqn:milp3}
        t_{nk} \geq 0, \ n \in N, \ k \in K,
    \end{equation}

    \begin{equation}\label{eqn:milp4}
        t_{nk} \leq d_{nk}T_{max}, \ n \in N, k \in K,
    \end{equation}

    \begin{equation}\label{eqn:milp5}
        \sum_{k \in K} d_{nk} = 1, \ n \in N,
    \end{equation}

    \begin{equation}\label{eqn:milp6}
        \sum_{n \in N} \sum_{k \in K} t_{nk}v_{pk} = s_p, \ p \in P,
    \end{equation}

    \begin{equation}\label{eqn:milp7}
        \sum_{k \in K} d_{nk} q_{pk} - \sum_{k \in K} d_{n-1,k} q_{pk} \leq y_{pn}, \ p \in P, \ n \in N,
    \end{equation}

    \begin{equation}\label{eqn:milp8}
        \sum_{n \in N} y_{pn} = 1, \ p \in P,
    \end{equation}

    \begin{equation}\label{eqn:milp9}
        a_{k_{1},k_{2}}d_{n_{1},k_{2}} (n_1 + 1) \leq a_{k_{1},k_{2}}(d_{n_{2},k_{1}} + (1 - d_{n_{2},k_{1}})e)n_2, \ k_{1}, k_{2} \in K, \ n_{1}, n_{2} \in N,
    \end{equation}

    \begin{equation}\label{eqn:milp10}
        d_{nk} \in \left\{0, 1\right\}, \ y_{pn} \in \left\{0, 1\right\}, \ p \in P, \ n \in N, \ k \in K.
    \end{equation}

    The objective function~(\ref{eqn:milp1_2}) defines the makespan criterion. Inequality~(\ref{eqn:milp3}) guarantees that the duration of execution of configuration ${k}$ at the event point ${n}$ is non-negative, and inequality~(\ref{eqn:milp4}) guarantees that the duration of execution of configuration ${k}$ will be zero only if this configuration is not executed at the event point ${n}$, otherwise it will be no more than ${T_{max}}$. Equality~(\ref{eqn:milp5}) means that one and only one configuration is performed at each event point, and equality~(\ref{eqn:milp6}) means that each job must be completed completely. Inequality~(\ref{eqn:milp7}) and equality~(\ref{eqn:milp8}) guarantee continuity of job. Inequality~(\ref{eqn:milp9}) sets a partial order between configurations. Expression~(\ref{eqn:milp10}) describes the range of the ${d_{nk}}$ and ${y_{pn}}$ variables.

\begin{proposition} \label{prop:num_ep}
There is an optimal solution to MILP model~(\ref{eqn:milp1_2})--(\ref{eqn:milp10}) using~${2m + 1}$ event points, which defines an optimal schedule in Formulation~F1.
\end{proposition}

    \textbf{Proof.} Note that each event point corresponds to a change of configurations, and a change only occurs when any job (or several jobs) has begun or has ended. Suppose that at each event point only one job begins or ends, then it is easy to see that in this case ${2m}$ event points are needed. We also take into account that we need a zero point of events, the point at which no configuration is performed. This implies that in the case when no two jobs start and end at the same time, the number of event points is ${2m + 1}$. In any other case, a smaller number of event points would be required. $\Box$

    Thus, in what follows we assign ${e: = 2m}$.

%    The solutions of the MILP model were found by CPLEX solver with using GAMS modeling system.

    It is worth noting that the solutions to MILP model~(\ref{eqn:milp1_2})--(\ref{eqn:milp10}) prodive the information about which configurations are performed at which event point, but do not contain the distribution of jobs to the cores. For scheduling the execution of jobs on the processor cores (as Formulation~F1 requires), the following Algorithm~2 is proposed which takes as input ${k_{1}, k_{2},\dots, k_{h}}$, configurations sorted by execution order, as well as their execution time ${l_{k_{1}}, l_{k_{2}}, ..., l_{k_{h}}}$ and the number of cores ${c}$. The algorithm returns the staring time ${u_{p}}$ and the completion time ${f_{p}}$ for each job ${p}$.

    \vspace{5mm}
    \textbf{Algorithm 2.} Jobs scheduling on the basis of MILP solution

    \textbf{Step 1.} For each job ${p}$ from ${k_{1}}$ assign a free core and set ${u_{p} := 0}$

    \textbf{Step 2.} For each ${k_{i}, i = 2, ..., h}$ do:

    \indent \indent \textbf{2.1} For each job ${p \in k_{i} \cap k_{i-1}},$ keep the same core.

    \indent \indent \textbf{2.2} For each job ${p \in k_{i-1} \backslash k_{i}}$ free the core on which job ${p}$ was performed, and set ${f_{p} := \sum_{j=1}^{i-1}l_{k_{i}}}.$

    \indent \indent \textbf{2.3} For each job ${p \in k_{i} \backslash k_{i-1}}$ assign a free core and set ${u_{p} := \sum_{j=1}^{i-1}l_{k_{i}}}.$

%    Thus, using the above algorithm, it is possible to schedule the execution of jobs on the processor cores by solving the % MILP model (1) - (10) in polynomial time.

    \section{Greedy Algorithm} \label{sec:greedy}

       In view of the fact that the problem is NP-hard, a constructive heuristic has been proposed for formulation~F2. In what follows, this heuristic is called the {\em greedy algorithm}, because it assigns jobs to all cores, not allowing them to stand idle, if possible. At each iteration, the algorithm selects a set of jobs (configuration) to perform, trying to select jobs so that when allocation the data bus between them, each job gets the most closest share of the data bus to the one it needs, but at the same time the maximum possible number of cores should be loaded. After selecting a configuration, the greedy algorithm determines the completion of which of the selected jobs will lead to switching the next configuration. To this end, firstly, the algorithm calculates what percentage of the data bus will be allocated to each job, and then, on the basis of these data, it determines the speed of processing the selected jobs.
To give a detailed description of the greedy algorithm, let us denote by ${k_{1}, k_{2}, ..., k_{i}, ...}$ the sequence of configurations generated by the greedy algorithm, and denote by ${duration_{i}}$ the duration of the configuration ${k_{i}}$. Then the algorithm can be written as follows.

    \vspace{5mm}
    \textbf{Algorithm 3. Greedy algorithm}

    \textbf{Step 0.} Put percentage of free data bus ${freePercent := 100\%}$ (the entire data bus is free); number of free cores ${freeCores := c}$ (all cores are free); ${i := 1}$; ${k_{i} := \emptyset}$; ${duration_{i} := 0}$; time remaining for job ${p}$ to completion under ideal conditions ${leftTime_{p} := s_{p}}$; a set of all jobs that have not started yet: ${jobs := \left\{1,..., m\right\}}$

    \textbf{Iteration} $i$. Repeat Steps 1--7:

    \textbf{Step 1.} While ${freePercent > 0}$ and ${freeCores}$ is not ${0}$, repeat 1.1.-1.2:

    \indent \indent \textbf{1.1.} Find an admissible (not started earlier and not forbidden by the partial order) job ${p \in jobs}$ for which the value ${\left| freePercent - b_{p} \right|}$ is minimal. In other words, find such valid job ${p \in jobs}$, which has the bus requirement closest to ${freePercent}$. If no such job is found, then go to \textbf{Step 3}.

    \indent \indent \textbf{1.2.} Put

    \indent \indent \indent ${freePercent := freePercent - b_{p}}$;

    \indent \indent \indent ${freeCores := freeCores - 1}$;

    \indent \indent \indent ${jobs := jobs - \left\{p\right\}}$;

    \indent \indent \indent ${k_{i} := k_{i} \cup \left\{ p \right\}}$.

    \textbf{Step 2.} While ${freeCores}$ is not ${0}$, repeat 2.1.-2.2:

    \indent \indent \textbf{2.1.} Find an admissible job ${p \in jobs}$ that has the lowest data bus consumption. If no valid job is found, then go to \textbf{Step 3.}

    \indent \indent \textbf{2.2.} Put

    \indent \indent \indent ${freeCores := freeCores - 1}$;

    \indent \indent \indent ${jobs := jobs - \left\{p\right\}}$;

    \indent \indent \indent ${k_{i} := k_{i} \cup \left\{ p \right\}}$.

    \textbf{Step 3.} If ${k_{i} = \emptyset}$, then go to \textbf{Step 8}. Otherwise, distribute the data bus capacity between the jobs according to Algorithm~1, which gives the value ${z_{pk_{i}}}$ -- allocated percentage of the data bus to job~${p}$ in configuration~${k_{i}}$.

    \textbf{Step 4.} Calculate processing speed ${v_{pk_{i}}}$ of all jobs ${p \in k_{i}}$ in configuration ${k_{i}}$:

    \indent \indent \indent ${v_{pk_{i}} := z_{pk_{i}} / b_{p}}$.

    \textbf{Step 5.} Determine which job will be fully completed first in the chosen configuration and set the duration of the configuration ${k_{i}}$ equal to the duration of this job in the configuration~${k_{i}}$:

    \indent \indent \indent ${duration_{i} := min_{p \in k_{i}} \left\{ leftTime_{p} / v_{pk_{i}} \right\} }$.

    \textbf{Step 6.} For all ${p \in k_{i}}$, for which ${leftTime_{p} / v_{pk_{i}}}$ is equal to ${duration_{i}}$, set ${leftTime_{p} := 0}$.

    \textbf{Step 7.} Put

    \indent \indent \indent ${freeCores := c}$;

    \indent \indent \indent ${freePercent := 100\%}$;

    \indent \indent \indent ${k_{i+1} := \emptyset}$.

    For all ${p \in k_{i}}$, for which ${leftTime_{p} /v_{pk_{i}}}$ is not equal to ${duration_{i}}$, put

    \indent \indent \indent ${leftTime_{p} := leftTime_{p} - duration_{i} v_{pk_{i}} / s_{p}}$;

    \indent \indent \indent ${k_{i+1} := k_{i+1} \cup \left\{ p \right\} }$;

    \indent \indent \indent ${i := i + 1}$;

    \indent \indent \indent ${freePercent := freePercent - b_{p}}$;

    \indent \indent \indent ${freeCores := freeCores - 1}$.

    \textbf{Step 8.} Distribute the jobs among the cores according to Algorithm~2.

    \vspace{5mm}

    It is not difficult to see that the greedy algorithm constructs a feasible schedule and may be implemented with time complexity~${O(n^2)}$.

    \section{Methods of Data Generation and Schedules Testing} \label{section:data_generation}

    All calculations described in Sections~\ref{section:data_generation} and~\ref{section:computational_experiment} were carried out on a computer with 16 GB of RAM and Intel Core i7-8565U 1.80 GHz CPU. The operating system used was Windows 10 version 1909. The number of threads used for calculations did not exceed the number of processor cores, so the impact of other processes and the operating system itself can be considered insignificant.
    Turbo Boost \cite{Intel_turbo_boost} and Hyper-threading \cite{Intel_hyper_threading} options were turned off in order to be sure that the CPU temperature and other uncontrolled factors do not influence the jobs processing times.
    All the programs described below were implemented in C++. For the computational experiment, the following procedures taken from the Intel MKL (Math Kernel Library) are used as jobs:

    \begin{itemize}
        \item copying a vector to another vector,
        \item calculation of the sum of magnitudes of the vector elements,
        \item calculation of a vector-scalar product and adding the result to a vector,
        \item calculation of the QR factorization of a matrix.
    \end{itemize}

    The choice of such procedures is due to the fact that they consume the data bus in different ways. The input data to the procedures has different sizes (for procedures with vectors, this is the vector length, for procedures with matrices, this is the matrix size). For vectors, the dimensions from 10 to 70~million elements were used, for matrices, the dimensions varied from 1000 to 1300. Such sizes are due to the requirement that the jobs data should not be kept in the processor's cache and their durations should not be too small
%(not less than one second),
(otherwise large measurement errors can occur)
%. Durations also
and they should not be too large
%(not more than 10 seconds), because
(otherwise the measurements will take too much CPU time).

    Input parameters for the generator:

    \begin{itemize}
        \item The number of jobs for which a schedule needs to be made. Values used: 4, 6, 7, 8, 10 (finding optimal solutions for 11 or more jobs takes about one hour, and for 13 and more jobs the generated model size is more than 26 Gb).
        \item Partial order to be generated for the jobs. Values used: (i)~with a trivial partial order (no dependencies between the jobs), (ii)~constructed at random (With probability~${0.5}$ we decide that a job ${p_{1}}$ should be performed after job~${p_{2}}$. To avoid cycles, only pairs of jobs ${ \left( p_{1}, p_{2} \right) }$ where ${ \left( p_{1} > p_{2} \right) }$ are considered.), (iii)~a binary tree, and (iv)~one-to-many-to-one.
        % (first one job, after it many jobs, after them again one).
        \item Number of cores. Values used: 2, 3, 4.
    \end{itemize}

%    All calculated data are saved to a file in the format of the GAMS modeling system.
%    It is worth noting that, for the convenience of testing, the execution time under ideal conditions in milliseconds was taken as the number of units of time required for each job to fully complete. Thanks to this, the answer is also obtained in milliseconds and the optimal solution found can easily be compared with the actual execution of job according to the schedule.

    \subsection{Calculation of the Data Bus Consumption}

    All data, except for the data bus consumption by each job, necessary for the greedy algorithm, was taken from the examples generated in Section~\ref{section:data_generation}: the number of time units needed for each job, the partial order, and the number of cores. To calculate the percentage of the data bus consumed by job~${p}$, we used a program that works as follows:
\begin{itemize}
\item    Job ${p}$ was started simultaneously in ${c}$ copies (${c}$ is equal to the number of cores on the computer used), after that the speed ${s_{p}^*}$ of the job was calculated as the execution time of job ${p}$ under ideal conditions, divided by the execution time of job~${p}$ along with ${ \left( c - 1 \right) }$ copies of the same job, then ${\frac{100\%}{s_{p}^{*} c_{i}}}$ is taken as the desired data bus consumption.

\item    If ${c}$ copies of job ${p}$ running simultaneously did not slow down each other, then this job was started with ${ \left( c - 1 \right) }$ copies of job ${g},$ which has the highest data bus consumption. In this case, the percentage of data bus consumption by job ${p}$ can be found as follows: ${100\% - \left( c - 1 \right) x}$, where ${x}$ is the percentage of the data bus required by job~${g},$ multiplied by the speed of~${g}$ in this configuration. If, in this case, no job has slowed down, then job~${p}$ in any configuration does not affect the speed of other jobs, therefore, the data bus consumption by job~${p}$ can be set equal to~${0}$.
\end{itemize}

    \subsection{Experimental Measurement of Makespan for the Constructed Schedules} \label{subsection:schedule_execution}

    Using the generator from Subsection~\ref{section:data_generation}, we calculated the real speed of jobs in various configurations. However, the greedy algorithm calculates these speeds based on data on the consumption of the data bus by each job according to formulation~F2. In order to understand how adequate the completion times are estimated in the MILP model using formulation~F1 and in the greedy algorithm using formulation~F2, and how the greedy solutions compare to the MILP solutions, a program code was written that simulated the execution of a given schedule on the processor cores.
%This program calculates the real execution time of the jobs according to the schedules built by the greedy algorithm and the CPLEX package, which allows us to compare the solutions obtained using the formulations F1 and F2 with each other.

In this testing program, threads are created in an amount equal to the number of cores of the simulated processor. Each thread is passed a job queue in the order in which they need to be executed. Before starting to perform the next job, the thread expects the completion of all previous jobs.
%To draw up job queues for each thread, an algorithm was suggested that first determines the order of the job, and then distributes the job into queues, given that in the process of executing the job, the core on which it is executed cannot change.
If the core should be idle between performing two jobs, then a fictitious job is added to the queue between the corresponding jobs.
%, which suspends the execution of the current thread by the number of time units equal to the idle time.

    \section{Computational Experiment} \label{section:computational_experiment}

\begin{figure}
        \begin{center}

    \includegraphics[height=6.6cm, width=\textwidth]{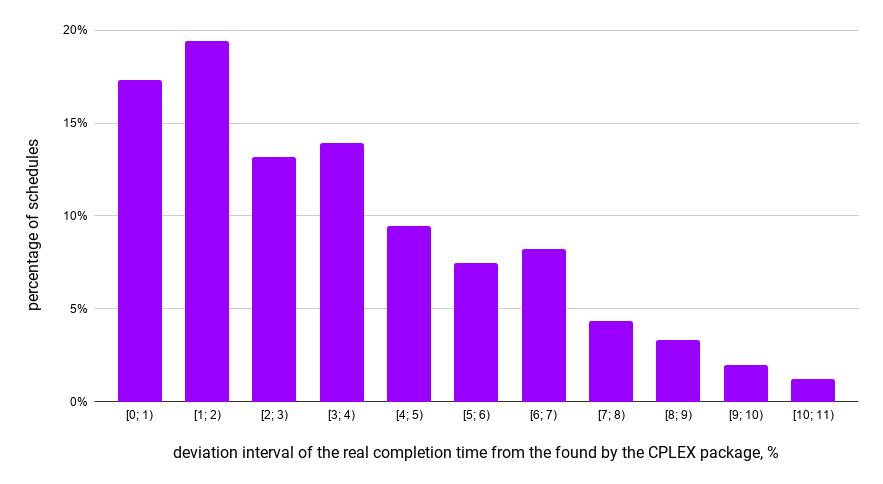}
    \caption{Histogram of relative deviation of the minimum completion time in formulation~F1 from the real completion time}
    \label{gams_fix_mode}
        \end{center}
\end{figure}

    Schedules constructed using GAMS modeling system with MILP model~(\ref{eqn:milp1_2})-(\ref{eqn:milp10}) were tested using the program described in Subsection~\ref{subsection:schedule_execution}. Fig.~\ref{gams_fix_mode} shows the histogram of relative deviation (in percentage) of the makespan calculated by the CPLEX package from the measured makespan.     In total, 964 schedules were tested, in all of them the deviation does not exceed 11\%, in 98\% of them it does not exceed 10\%, and in 73\% it does not exceed 5\%.

\begin{figure}
    \begin{center}       
     \includegraphics[height=8cm,width=\textwidth]{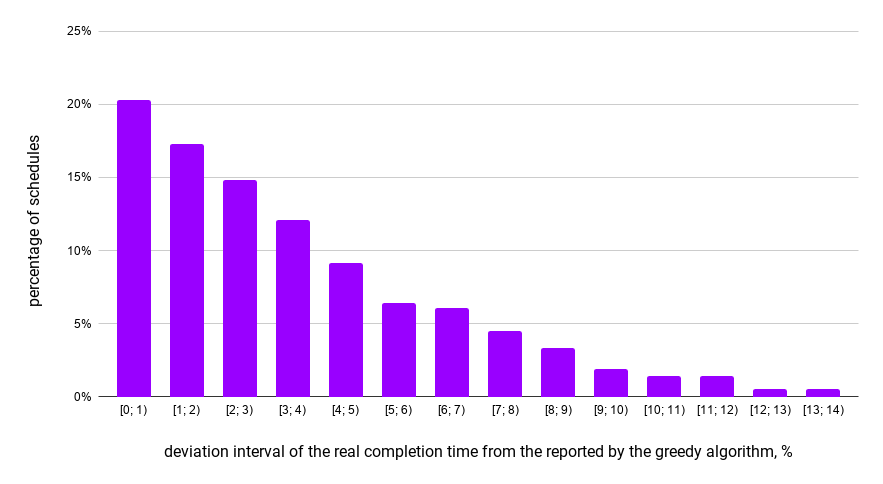}
        \caption{Histogram of relative deviation of completion time reported by the greedy algorithm from the real completion time}
        \label{ga_fix_mode}
    \end{center}
    \end{figure}

    Schedules constructed by the greedy algorithm were also tested using the program described in Subsection~\ref{subsection:schedule_execution}. Fig.~\ref{ga_fix_mode} shows a histogram of relative deviation (in percentage) of completion time reported by the greedy algorithm from the real completion time.

    In total 984 schedules with different number of jobs, different partial orders, and different number of cores were tested. In most of the cases (95\%), the makespan evaluation computed in greedy algorithm differs from that obtained in the experiment by no more than 10\%, and in 73\% of cases by at most~5\%. In 100\% of cases the deviation does not exceed 14\%. Such results show a fairly high accuracy of evaluation of the jobs processing time in the greedy algorithm.

\begin{figure}
    \begin{center}
        \includegraphics[width=10cm]{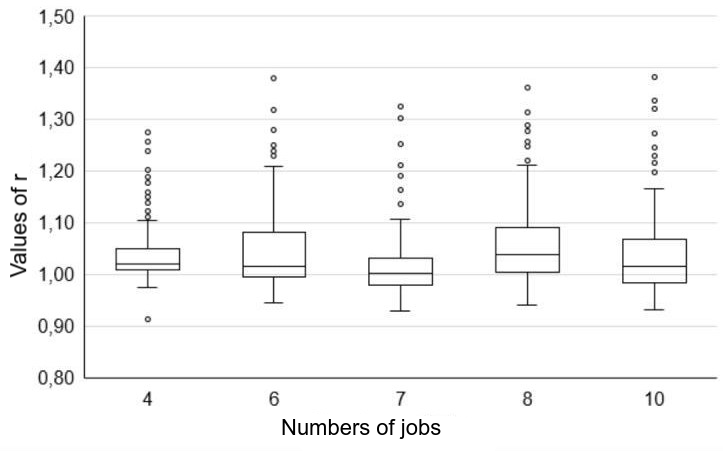}
        \caption{Ratio of the real measured makespan for greedy schedules in formulation~F2 to the measured makespan of optimal schedules in formulation~F1}
        \label{fix_mode_results}
    \end{center}
    \end{figure}

    Let us denote by ${r: = \frac{ga\_f2_{real}}{opt\_f1_{real}}}$ the ratio of the real measured makespan of the greedy schedules (${ga\_f2_{real}}$) in formulation~F2 to the real measured makespan of the optimal schedules (${opt\_f1_{real}}$) in formulation~F1. Fig. \ref{fix_mode_results} shows a box-plot diagram of~$r$ ratio for different numbers of jobs.

    For each number of jobs, ${192}$ schedules with different partial orders and number of cores were tested. It can be noted that the median ratio~$r$ for each number of jobs is close to~${1.05}$, which allows us to conclude that the greedy algorithm is highly accurate. It is also worth noting that even in the worst cases, the makespan of greedy schedule exceeds the mistaken found by the MILP model in formulation~F1 not more than by a factor of~${1.4}$. In Fig.~\ref{fix_mode_results}, one can also see that in some cases the solutions of the greedy algorithm in real life turn out to be faster than the optimal solutions, however, the difference does not exceed ${10\%}$ and may be related to the error in calculating the input data of the problem and in testing of the obtained solutions.

    \begin{table}
        \begin{center}
            \begin{tabular}{ | r | r | r | r | r | r | }
                \hline
                \textbf{} & \textbf{4 jobs} & \textbf{6 jobs} & \textbf{7 jobs} & \textbf{8 jobs} & \textbf{10 jobs} \\ \hline
                No ordering & $0.4$ sec & $4.3$ min & $13$ min & $16$ min & $15.5$ min\\
                One to many to one & $0.2$ sec & $3$ sec & $26$ sec & $6.3$ min & $14.8$ min\\
                Random order & $0.2$ sec & $3.6$ sec & $18$ sec & $32$ sec & $3.6$ min \\
                Bitree order & $0.2$ sec & $4$ sec & $1.5$ min & $7.2$ min & $16$ min \\
                \hline
            \end{tabular}
        \end{center}
        \caption{Average CPU time of the CPLEX package}
        \label{gams_time_fix_mode_table}
    \end{table}

    \begin{table}
        \begin{center}
            \begin{tabular}{ | r | r | r | r | r | r | }
                \hline
                \textbf{} & \textbf{4 jobs} & \textbf{6 jobs} & \textbf{7 jobs} & \textbf{8 jobs} & \textbf{10 jobs} \\ \hline
                No ordering & $2.5$ \si{\micro\second} & $3.9$ \si{\micro\second} & $4.8$ \si{\micro\second} & $6.4$ \si{\micro\second} & $8.1$ \si{\micro\second}\\
                One-to-many-to-one & $2.5$ \si{\micro\second} & $4.6$ \si{\micro\second} & $5.8$ \si{\micro\second} & $6.9$ \si{\micro\second} & $9.6$ \si{\micro\second}\\
                Random order & $2.6$ \si{\micro\second} & $4.4$ \si{\micro\second} & $6.2$ \si{\micro\second} & $7.7$ \si{\micro\second} & $11.4$ \si{\micro\second} \\
                Bitree order & $2.8$ \si{\micro\second} & $4.5$ \si{\micro\second} & $6$ \si{\micro\second} & $6.9$ \si{\micro\second} & $9.5$ \si{\micro\second} \\
                \hline
            \end{tabular}
        \end{center}
        \caption{Average CPU time of the greedy algorithm}
        \label{ga_time_fix_mode_table}
    \end{table}

    Tables~\ref{ga_time_fix_mode_table} and~\ref{gams_time_fix_mode_table} show the CPU time of the greedy algorithm (Table~\ref{ga_time_fix_mode_table}) and the CPU time of the CPLEX package (Table~\ref{gams_time_fix_mode_table}) for different types of partial order and different numbers of jobs.
    The CPLEX package most quickly finds solutions for jobs with random partial order, since this type of partial order is usually more constraining than others, and most slowly for the trivial partial order. The greedy algorithm, on the contrary, works faster with trivial partial order, and slower for the random partial order. Still for all types of partial order and for any number of jobs it is much faster than CPLEX.

    \section{Conclusions} \label{section:conclusions}

    In the paper, the problem of multi-core processor scheduling was analyzed taking into account the bandwidth limitations of the data bus. Two problem formulations are suggested. A mixed integer linear programming model is proposed for the first problem formulation and the MILP solutions were found by CPLEX solver. A greedy algorithm for approximate solving the problem is proposed for the second problem formulation.

The schedules found by the CPLEX package and the greedy algorithm were tested using a program simulating the execution of jobs on the processor cores.
%The results showed that the makespan indicated in the schedule deviates from the real time of execution in most cases by no more than 10\%, both for solutions of the greedy algorithm and for solutions of the CPLEX package.
The greedy algorithm has only a quadratic running time and a fairly high accuracy: a real-life testing showed that in 83\% of the cases the makespan of a greedy schedule deviated from the optimal solution of MILP model less than by~10\%, and in 60\% of the cases the deviation was less than by~5\%. We can conclude that the proposed algorithm of calculation of data bus consumption for a given configuration and the method of calculating the speed of jobs based on these data are close to what happens in real life.

    \section*{Acknowledgment}

The work was funded by project~0314-2019-0019 of Russian Academy of Sciences (the Program of basic research~ I.5).

    \bibliographystyle{splncs03}

\begin{thebibliography}{99}
        \bibitem{Althaus18} Althaus, E., Brinkmann, A., Kling, P. et al.: Scheduling shared continuous resources on many-cores. J Sched \textbf{21}, 77–-92 (2018)

        \bibitem{Brauner} Blazewicz, J., Brauner, N., Finke, G.: Scheduling with discrete resource constraints. In: J.Y-T. Leung (ed.) Handbook of Scheduling, pp. 23-1 -- 23-18. CRC Press, Boca Raton (2004)

        \bibitem{GJ} Garey, M.R., Johnson, D.S.: Computers and intractability. A guide to the theory of NP-completeness. W.H.~Freeman and Company, San Francisco (1979)

        \bibitem{Ierapetritou} Ierapetritou M.G., Floudas C.A.: Effective continuous-time formulation for short-term scheduling. 1. Multipurpose Batch Processes \textbf{37}(11), 4341--4359 (1998)

        \bibitem{Intel_hyper_threading} Intel Hyper-Threading Technology. \url{https://www.intel.com/content/www/us/en/architecture-and-technology/hyper-threading/hyper-threading-technology.html}

        \bibitem{Intel_turbo_boost} Intel Turbo Boost Technology 2.0. \url{https://www.intel.ru/content/www/us/en/architecture-and-technology/turbo-boost/turbo-boost-technology.html}

        \bibitem{Jiang} Jiang, Y., Shen, X., Chen, J., Tripathi, R.: Analysis and approximation of optimal co-scheduling on chip multiprocessors. In: Proc. of the 17th Int. Conf. on Parallel Architectures and Compilation Techniques (PACT'08), pp.~220--229 (2008)

        \bibitem{Jozefowska} Jozefowska, J., Weglarz, J.: Scheduling with resource constraints – continuous resources. In J. Y.-T. Leung (Ed.), Handbook of Scheduling, pp. 24-1 -- 24-15. CRC Press, Boca Raton (2004)

        \bibitem{Liu} Liu M., Chu F., He J., Yang D., Chu C.: Coke production scheduling problem: A parallel machine scheduling with batch preprocessings and location-dependent processing times. Computers and Operations Research \textbf{104}, 37--48 (2019)

        \bibitem{Merkel} Merkel, A., Stoess, J., Bellosa, F.: Resource-conscious scheduling for energy efficiency on multicore processors. In Proceedings of the 5th European Conference on Computer Systems (EuroSys'10), pp. 153--166 (2010)

        \bibitem{Nesbit06} Nesbit, K.J., Aggarwal, N., Laudon, J., Smith, J.E.: Fair Queuing Memory Systems. In Proc. of 39th Annual IEEE/ACM International Symposium on Microarchitecture (MICRO'06), pp. 208--222 (2006)

        \bibitem{Servakh} Servakh, V. V.: Effectively solvable case of the production scheduling problem with renewable resources. Diskretn. Anal. Issled. Oper., Ser. 2 \textbf{7}(1), 75--82 (2000) (in Russian)

        \bibitem{Shabtay} Shabtay, D., Kaspi, M.: Parallel machine scheduling with a convex resource consumption function. Eur. J. Oper. Res. \textbf{173}(1), 92--107 (2006)

        \bibitem{SM90} Shahrokhi F., Matula D. W.: The maximum concurrent flow problem. J. ACM \textbf{37}(2), 318--334 (1990)

        \bibitem{Strusevich} Shioura, A., Shakhlevich, N.V., Strusevich, V.A.: Preemptive models of scheduling with controllable processing times and of scheduling with imprecise computation: A review of solution approaches. 	Eur. J. Oper. Res. \textbf{266}(3), 795--818 (2018). 
        %doi: 10.1016/j.ejor.2017.08.034

        \bibitem{Tian} Tian, K., Jiang, Y., Shen, X., Mao, W.: Optimal Co-Scheduling to Minimize Makespan on Chip Multiprocessors. 
        %In: Cirne, W., Desai, N., Frachtenberg, E., Schwiegelshohn, U. (eds) 
        In: Proc. of Job Scheduling Strategies for Parallel Processing (JSSPP'2012), Lecture Notes in Computer Science, vol.~7698. Springer, Berlin, Heidelberg (2013)

        \bibitem{Xiao} Xiao, Z., Chen, L., Wang, B., Du, J., Li, K.: Novel fairness-aware co-scheduling for shared cache contention game on chip multiprocessors. Inf. Sci.
%        Information Sciences 
\textbf{526}, 68--85 (2020)

        \bibitem{Zhuravlev} Zhuravlev, S., Blagodurov, S., Fedorova, A.: Addressing shared resource contention in multicore processors via scheduling. In Proc. of the 15th Int. Conf. on Architectural Support for Programming Languages and Operating Systems (ASPLOS'10), pp. 129--142 (2010)


%        \bibitem{Kovalenko_article} Kovalenko, Yu. V.: On the calendar planning problem with renewable resource. Automation and Remote Control  \textbf{73}, 1046--1055 (2012)
%        \bibitem{Gams} The General Algebraic Modeling System (GAMS), \url{https://www.gams.com/latest/docs/index.html}. Last accessed 5 Jul 2020
%        \bibitem{Gawiejnowicz} Gawiejnowicz S, Kononov A.: Isomorphic scheduling problems. Annals of Operations Research \textbf{213}(1), 131--145 (2014)
%        \bibitem{Hager} Hager G., Treibig J., Habich J., Wellein G.: Exploring performance and power properties of modern multicore chips via simple machine models, \url{https://arxiv.org/pdf/1208.2908.pdf}. Last accessed 5 Jul 2020
%        \bibitem{Intel_mkl} Developer Reference for Intel Math Kernel Library (Intel MKL), \url{https://software.intel.com/content/www/us/en/develop/documentation/mkl-developer-reference-c/top.html}. Last accessed 5 Jul 2020
%        \bibitem{Ilic} Ilic A., Pratas F., Sousa L.: Cache-aware roofline model: Upgrading the loft, \url{https://www.inesc-id.pt/ficheiros/publicacoes/9068.pdf}. Last accessed 5 Jul 2020
%        \bibitem{Jansen} Jansen K., Porkolab L.: Preemptive parallel task scheduling in O(n) + Poly(m) time. Lee D.T., Teng S.H. (eds.) ISAAC 2000. LNCS, 398--409 (2000).
%        \bibitem{Leiserson} Leiserson C. E. A Minicourse on Dynamic Multithreaded Algorithms, \url{https://acikders.tuba.gov.tr/file.php/132/Readings/L22_dyn_multi_alg.pdf}. Last accessed 5 Jul 2020
    \end{thebibliography}

\end{document}